# From design to tape-out in SCL 180nm CMOS integrated circuit fabrication technology

Joydeep Basu

Joydeep Basu is with the Department of Electronics & Electrical Communication Engineering (E&ECE) and Advanced VLSI Design Laboratory (AVDL), Indian Institute of Technology (IIT) Kharagpur, Kharagpur 721302, India.

**ABSTRACT**
Although India has achieved considerable capability in electronic chip design, but developing the infrastructure for capital-intensive semiconductor fabrication remains a challenge. The rising domestic and global demand for electronics products, the need of enhancing the country's high-technology talent pool, employment generation, and national security concerns dictates the Indian Government's heightened efforts in promoting electronics hardware manufacturing in the country. A recent milestone in this regard is the setting up of 180nm CMOS fabrication facility at SCL, Chandigarh. The Multi Project Wafer runs of this indigenous foundry promises to be a relatively cost-effective option for Indian academic and R&D institutions in realizing their designed VLSI circuits. Written from the perspective of an Analog VLSI designer, this tutorial paper strives to provide all the requisite information and guidance that might be required in order to prepare chip designs for submission to SCL for fabrication.

## 1. INTRODUCTION

Electronics industry is one of the world's leading and fastest growing industrial sectors. However, the sluggish production of electronic products in India does not justify the country's huge and ever-increasing demand for such goods [1]–[3]. To address this gap, the Government of India is increasing its focus on the Electronics System Design and Manufacturing (ESDM) sector with an aim to transform the country from a largely import and consumption driven market to one with manufacturing capability. The Govt. is attributing high priority to electronics hardware manufacturing in the country, and strives to achieve "net zero import" as notified in its national policy on electronics (NPE) [4]. This is also foundational for success of the *Make in India* and *Digital India* initiatives of the Govt. The electronics sector is also poised to be a key driver for employment growth in India. The nation has already started witnessing decent traction in this regard with increased assembly activities across consumer products such as mobile phones. To advance the indigenous manufacturing of electronics for both domestic and international markets, the Indian Govt. is incentivizing Foreign Direct Investment and chip design start-ups in the ESDM sector through its various schemes [1], [4]. Emphasis is also placed on skill/knowledge development, and in upgrading the related infrastructure in research and academic institutions of the country.

Semi-Conductor Laboratory (SCL) near Chandigarh, India under the Department of Space of the Govt. of India, is a premier semiconductor research & development (R&D) unit of the country. SCL is at the forefront of the Govt. of India's effort towards developing a strong microelectronics base in India and enhancing capability in Very Large Scale Integrated (VLSI) circuit domain. The CMOS fabrication facility of SCL got upgraded a few years back to 180nm technology adapted from Tower Semiconductor of Israel [5]. This 8-inch wafer foundry is capable of fabricating integrated circuits (IC) for analog, mixed-signal, and digital applications; and is the country's sole establishment that can boast of such a facility.

Few salient features of SCL's CMOS process are provided below for the information of the reader [5], [6]:
- 1.8V core CMOS
- 1.8V or 3.3V I/O library cells
- Single Poly and 4 to 6 metal layers
- High $V_T$ MOSFETs
- Deep N-well (i.e., isolated p-wells) for noise isolation
- Metal-insulator-metal (MIM) capacitor
- High resistivity polysilicon resistor
- Selective Cobalt salicidation
- Ti/TiN/AlCu/Ti/TiN metal interconnects
- 1.8V standard cells for logic design

SCL offers opportunity to the academia to participate and contribute to research related to microelectronics and VLSI. This is implemented through the granting of funded research projects (e.g., the RESPOND programme sponsored by ISRO, which is targeted towards space related applications [7]) or, by the signing of memorandum of understanding (MoU) with various institutes/universities for research collaboration. A main appeal for the Indian academia in such engagement is the scope of getting their VLSI circuit designs fabricated from the in-house foundry of SCL. This is also imperative from the India Govt.'s perspective of supporting 3000 additional PhDs in ESDM and IT sectors as part of the NPE that emphasizes on significantly increasing the availability of skilled manpower, and in developing an ecosystem of research, development and IP creation in this knowledge intensive domain [8]. This is also pertinent for success of initiatives like the Special Manpower Development Programme (SMDP) that is targeted towards strengthening the research base and generation of trained human capital in the area of chip design [9]. The programme encompasses 60 academic and R&D institutions pan India (like the IITs, IISc, NITs, CEERI, etc.) and stresses on the realization of working prototypes of System-on-Chip (SoC). To reach such an end goal, a concerted and coordinated interaction between the academia and governmental firms like SCL is very vital, and can indeed be a defining step towards achieving India's much awaited self-reliance in the high-tech semiconductor design and manufacturing segment.

In view of the need of the academia to efficiently utilize the CMOS fabrication facility of SCL, this tutorial paper aims at providing all the requisite information and guidance that might be required in this direction. The paper as such is written from the perspective of an Analog VLSI designer, mainly utilizing CAD tools like *Cadence Virtuoso*, *Cadence Spectre* [10], and *Mentor Graphics Calibre* [11]. Nonetheless, the paper should be helpful for Digital designers as well. Various aspects right from transistor level design to tape-out in SCL's 180nm PDK (particularly issues like post-layout simulation, LVS, I/O ring, dummy metal fill, full chip DRC, GDS generation, finding check-sum, etc.) have been discussed. Such information is usually not available readily leading to substantial loss of time and effort on part of both the academic designer and the concerned liaison from SCL. The paper strives to bridge this information gap by building on the knowledge and experience gained during the course of chip design and submission to SCL while being supported by SCL engineers during the process.

## 2. CIRCUIT SCHEMATIC SIMULATION

### 2.1 Working Directory Structure

Details of basic environment and tool (e.g., *Cadence Virtuoso, Synopsys HSPICE, Mentor Graphics Calibre, etc.*) setup as required for designing using SCL PDK are not provided here, as it is required to be done by a person experienced in setting up of such tools (e.g., the lab admin) in a Linux environment.

Once the basic setup is ready, the user (designer) will be having a working directory in his/her terminal machine. It will have a structure as exemplified in Fig. 1. Description of the working directory items are as follows.
(i) The *cdl_netlist* directory contains the *CDL* netlists that will be required for LVS. It also contains *scale.cdl* file which is used during LVS and PEX run.
(ii) *physical_verification* contains four separate directories for DRC, Antenna, LVS, and PEX runs using Calibre.

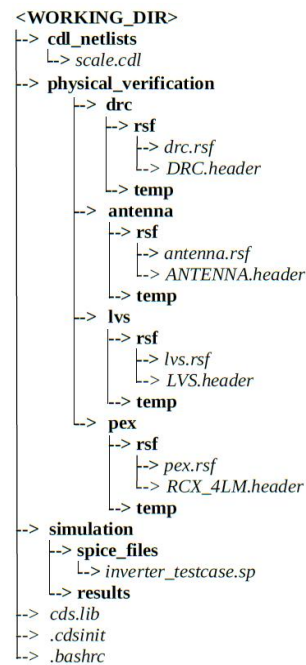

**Figure 1. Example working directory structure for SCL PDK usage.**

(iii) The *cds.lib* file contains the following:
- SCL 0.18um PDK library (*ts018_scl_prim*)
- 1.8V I/O library (*CIO150*)
- 3.3V I/O library (*CIO250*)
- HSPICE compatible analogLib (*analogLib_hspice*)

(iv) There can also be directories like *spice_files* and *results*. *spice_files* can contain the SPICE netlists used to run post-layout simulation using HSPICE simulator. Further, *results* directory can be used to store the simulation output files.

### 2.2 Circuit Simulation

It is assumed that the reader has adequate familiarity with simulations in *Cadence Virtuoso Analog Design Environment* (ADE), and so, this paper is not meant to be a comprehensive guide towards that. To perform simulation using SCL PDK, launch Virtuoso. From CIW window, click *Tools* → *Library Manager*. From the *Library Manager* window, click File → New → Library. In *Library section* of the ensuing form, enter a desired name in *Name*, and click *OK*. In the next dialog, select "Attach to an existing technology library." Following this, an additional dialog pops up where you should select "ts018_scl_prim" as the *Technology Library*, and click *OK*.

Now, click on the name of the newly created library in the Library Manager. Then, click *File* → *New* → *Cell View*. In the *New File* dialog that opens, provide the desired *Cell* name, and *Type* should be selected as "schematic". Once you click *OK*, the Virtuoso schematic editor opens. Here, compose the required transistor level schematic using devices/components instantiated from *ts018_scl_prim* library (e.g., nmos_18, pmos_18, etc.), analogLib, etc. Once done, click *File* → *Check and save* to save the schematic.

To simulate the schematic, the user has the option to either use HSPICE integrated into ADE [12], or the intrinsic Spectre simulator of Cadence [13]. As usage of the latter is more common in academia (and more convenient too from within ADE GUI) than HSPICE, this paper documents the usage of Spectre for all simulations.

After the schematic is completed, click *Launch → ADE-L*. Here, ensure that *Simulator* option is set to "spectre" (by using *Setup → Simulator*). Next, click on *Setup → Model Libraries*. Enter the path of the model file (an example being shown below), and choose appropriate *Section* from the drop-down list. E.g., select "tt_18" for typical 1.8V MOSFETs.

```
~/scl_pdk/design_kit/models/hspice/ts18sl_scl.lib
```

Separate model file sections are required to be used if other device types, like resistors (e.g., *res2t_typ*), capacitors (e.g., *mimcap_typ*), etc. are present in the schematic.

In SCL PDK, dimensions of the devices are assumed to be in micrometer scale. Thus, in ADE, ensure that the parameter *Scale* is set to 1e-6 (using *Simulation → Options → Analog → Component*). With these settings, you are ready to perform simulations in ADE.

## 3. PHYSICAL DESIGN

After schematic freeze, layout design is to be done followed by related DRC, LVS, etc. checking steps, as delineated below. SCL recommended die sizes are 2mm×2mm and 5mm×5mm.

### 3.1 Design of Layout

From the schematic editor window, launch *Layout-XL* in order to create the corresponding layout. The Layout Grid of layout editor must be set to an integer multiple of 0.005µm. Thus, using *Options → Display* in Layout-XL window, set *X Snap Spacing* and *Y Snap Spacing* to 0.005 (i.e., 5nm).

Now, prepare the layout design as desired. Some common layers that may be needed for drawing are WN, GC, ACTIVE, XP, XN, M1, M2, M3, TOP_M, CS, V2, V3, TOP_V (all with *purpose* "drw"). Corresponding description and DRC rules are available in [6]. To instantiate vias, click *Create → Via*, and in the *Via Definition* field, select as required. For using guard ring in the layout, select the area to be enclosed, and then click *Create → MPP Guard Ring*. In the resulting dialog, select the preferred option in *Guard Ring Template* field, e.g., choosing "PWELL Tap" or "PWELL RING" will enclose the selected area with a P-well guard ring. Then click *Apply* and *Hide*.

Remember to place matching labels (using suitable layer with *purpose* "lbl") for each corresponding pin as used in the schematic. Note: use layer WN (of *purpose* "vss") if you have multiple Grounds (e.g., Digital Ground and Analog Ground) in the layout to avoid LVS error (as electrically, both of these Grounds are connected to the same bulk).

### 3.2 Design Rule Check

To run a Design Rule Check (DRC) on the designed layout, use *Calibre* menu in Virtuoso layout editor window (for this, the Calibre-Virtuoso interface should be present installed). Click *Calibre → Run DRC* from the layout window. Go to *File→Load Runset* and browse to the *drc.rsf* file. Then, click on *Run DRC*. Rectify layout errors (if any) as pointed by DRC.

### 3.3 Antenna Rule Check

To check for antenna errors, invoke *Calibre → Run DRC* from the layout editor. Go to *File→Load Runset* and browse to the *antenna.rsf* file. Click on *Run DRC*, and rectify errors if any.

### 3.4 Layout Versus Schematic

In order to check the electrical propriety of the prepared layout with respect to its schematic, run Layout Versus Schematic (LVS) using the following steps.
(i) Create *CDL netlist* of the schematic using CIW→File→ Export→CDL. In the window that appears, click on the button *Library Browser* and browse to the desired schematic. In the *Output* section of the window, specify the *Run Directory* as the *cdl_netlist* directory in the <WORKING_DIR>. Also, specify the *Output CDL Netlist File* as the name of the cell. An example screenshot is shown in Fig. 2 for convenience.
(ii) Launch *Calibre → Run LVS* from the layout editor.
(iii) Go to *File→Load Runset* and browse to the *lvs.rsf* file.
(iv) In the *Inputs* section of the LVS window, select the *Netlist*

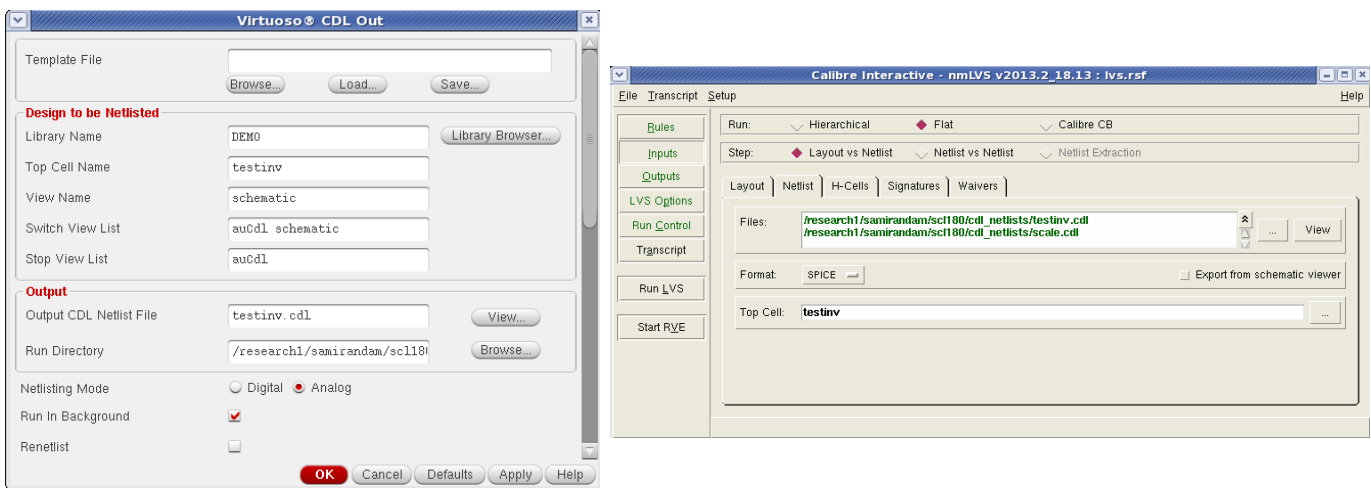

**Figure 2. Screenshots showing setup as required for creating the CDL netlist (left) and for running LVS using Calibre (right).**

tab. In the *Files* field, browse to your *cdl_netlist* directory and select the CDL file as generated before, and also the *scale.cdl* file, and add these in the list of files (as shown in Fig. 2).
(v) You can now click on *Run LVS*, and clean errors if any.

## 4. POST-LAYOUT SIMULATION

Steps for performing post-layout simulation on the extracted circuit netlist as obtained from the layout are discussed here.

### 4.1 Parasitics Extraction

Generate the parasitics extracted netlist using following steps.
(i) Launch *Calibre→Run PEX* from the layout editor window.
(ii) Go to *File→Load Runset* and browse to the *pex.rsf* file.
(iii) In the *Inputs* section of the PEX window, select the *Netlist* tab. In the *Files* field, browse to your *cdl_netlist* directory and select the CDL file (as generated before for the particular cell) and the *scale.cdl* file, and add these in the list of files.
(iv) Go to the *Outputs* section of the PEX window and choose *Format* as either *HSPICE* (if you wish to use HSPICE for post-layout simulation) or *SPECTRE* (if you want Spectre); and change the *File* name to <cell_name.sp> (for HSPICE) or to <cell_name.pex.netlist > (for Spectre).
(v) Go to *PEX Options* section and select the *LVS Options* tab. Here, enter VDD in the field *Power nets* and GND in the field *Ground nets*.
(vi) Following this, click on *Run PEX*. This will generate a parasitics extracted file named <cell_name.pex.netlist>.

### 4.2 Lone Simulation of a PEX Netlist

Post-layout simulation can be done with HSPICE in command-mode. For this, the user should write an appropriate SPICE testbench netlist. This should include reference to the PEX generated netlist. For reasons mentioned before, post-layout simulation with Spectre has been detailed here instead of HSPICE. Its various consecutive steps are mentioned below.

The netlist created by PEX is a file like <cell.pex.netlist>. Is it possible to associate this netlist to the corresponding sub-circuit symbol used in a bigger circuit in Virtuoso schematic editor. For this, you need to create a cell with a CDF parameter "model" which will point to the netlist that you want to use.

#### 4.2.1 CDF Edit

Go to Cadence CIW window, and click *Tools→CDF→Edit*. Select *Base* in "CDF layer". Then, choose the appropriate library name and cell name. Click on "Component Parameter" section in the CDF form, and write "model" in both the *Name* and *Prompt* columns. And select "string" in *Type* section. Also, set "yes" in *Parse as CEL* field. Keep the other fields at default values and click on *Apply*. An illustrative screenshot is given in Fig. 3.

Next, go to *Simulation Information* section and select "By Simulator" option. In the drop-down menu, select "spectre" and write "model" in *otherParameters* box. In "termOrder" field, write down the pin names in the proper order as in the corresponding PEX netlist file. Click on the button *OK*. These settings are depicted in Fig. 3. **Note:** The pin names entered here should match with that in the corresponding symbol (in case those names differ with the netlist). Also, the order of pins in the generated PEX netlist might change if the layout is re-extracted. Then, the order should be changed accordingly in CDF as well.

#### 4.2.2 Spectre View Generation of Cell

From Virtuoso Library Manager, select the *symbol* view of the cell (corresponding to the layout for which we need to do post-layout simulation), right click on it, and select *copy* option. In the "To" section of the ensuing *Copy View* window, change the *view* field from "symbol" to "spectre", and click *OK*. So, a *spectre* view is created that appears exactly like the symbol.

#### 4.2.3 Instantiate PEX Netlist into Schematic

Create a testbench schematic by instantiating the *symbol* (or alternatively, the *spectre* view) of the cell (whose layout PEX netlist is to be simulated). Now select the symbol and press "Q". Go to *CDF parameter* section in "Edit Object Properties" window and write the name of the corresponding cell in the "model" field.

Then, invoke ADE where the *Simulator* should be set to *Spectre*. In *Setup→Model libraries* form, provide the path to the SCL model files (as before) along with the path to the PEX netlist file. Go to *Simulation→Option→Analog→Component*, and set the "scale" field to 1 (as all the dimensions in the extracted PEX netlist are actual values and not scaled ones).

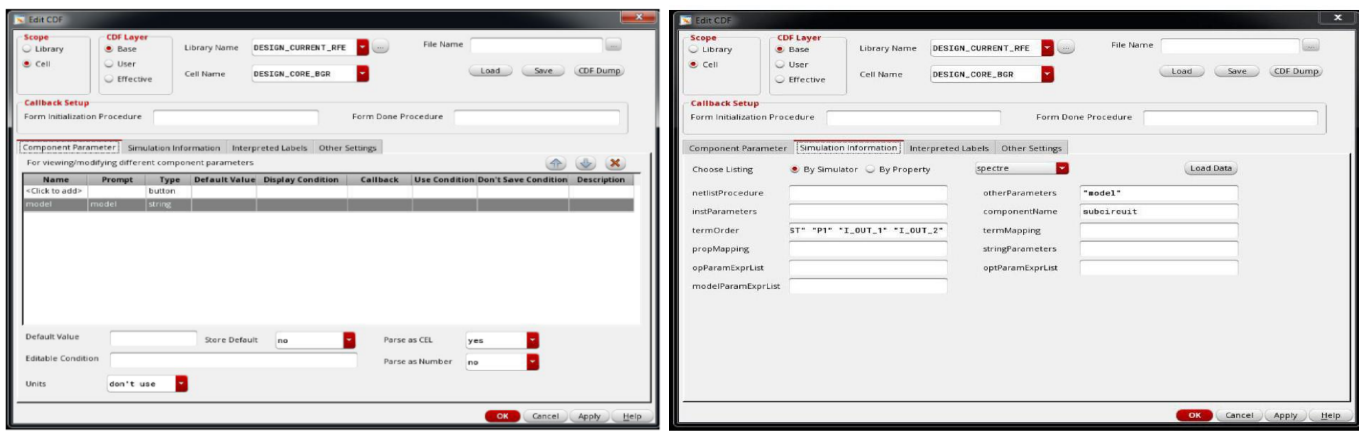

**Figure 3. Example screenshots illustrating the CDF editing procedure.**

With these settings, you are ready to run simulations. Mind that this method holds for multiple instantiations (sub-circuits) of PEX netlists in the testbench, but, would not work if some schematic level symbol is included in the testbench (due to mismatch in requirement of the "scale" field setting).

*Note:* If you want to switch between the schematic and the "spectre" (i.e., netlist) views, use the *Switch view list* in *Setup→Environment* in ADE and put "schematic" ahead of "spectre". You also need to remove the previously included sub-circuit netlist in *Model Libraries*; and change the "scale" value to the one as discussed above.

### 4.3 Co-Simulation of PEX Netlists with Schematics

Here, besides different constituent sub-circuits, the testbench schematic also includes a sub-circuit defined by a PEX netlist (associated using "spectre" view as discussed). Now, the issue is that the schematic sub-circuits require scale=1e-6; but, the sub-circuit defined in the netlist doesn't require this scale.

In this scenario, the simulation can be performed using the *MTS* (Multi Technology Simulation) mechanism in ADE-XL. The following steps are required to be followed for this. Let the netlist defined sub-circuit be named "Comparator_ver2a".

(i) At first, create a wrapper i.e., a schematic (say, named "Comparator_ver2a_wrapper") which instantiates symbol of "Comparator_ver2a", and also has the same symbol view as "Comparator_ver2a" - simply wire through all the connections: (also, do enter the corresponding cell name in the field "model" as shown below; the field "model" actually appears due to the prior CDF editing step).

(ii) Use the "Comparator_ver2a_wrapper" symbol (in place of "Comparator_ver2a symbol") in the main testbench schematic. Then, use the Hierarchy Editor to create a *config* view for your design. As depicted in Fig. 4, select the view to use as *spectre* for the "Comparator_ver2a" block in the Hierarchy Editor.

(iii) Launch ADE-XL from this config schematic window. Define a test by clicking on "Click to add test" - select *View Name* as *config* in *Choosing Design* dialog, which opens the ADE-XL Test Editor window. You may need to set Setup→simulator→spectre. Then, define the simulations and outputs required. After this, you can close the ADE-XL Test Editor.

(iv) Right Mouse click over the test name in ADE-XL to pick *Simulator*, and turn on the MTS checkbox there. Then, do Right Mouse click over the test name→*MTS Options*. In the MTS setup, enable "Comparator_ver2a_wrapper" as the MTS block, and add the model file (the netlist path) for "Comparator_ver2a" and scale=1.0 to it.

(v) Set the default scale in Simulation→Options→Analog to 1e-6 as usual.

Then you can run the simulation from ADE-XL. *Note:* To change back to the schematic view (instead of the netlist) for "Comparator_ver2a": In the Hierarchy Editor, change view to use to *schematic* for the "Comparator_ver2a" block, and click on *Recompute the hierarchy* button. Also, in the *MTS Options* window, uncheck the model file (i.e., the path of the netlist) for "Comparator_ver2a_wrapper" (so the netlist would not be used), and set its scale to 1e-6. Then you can run. Unchecking *MTS_BLOCK* for "Comparator_ver2a", or disabling MTS fully will also work.

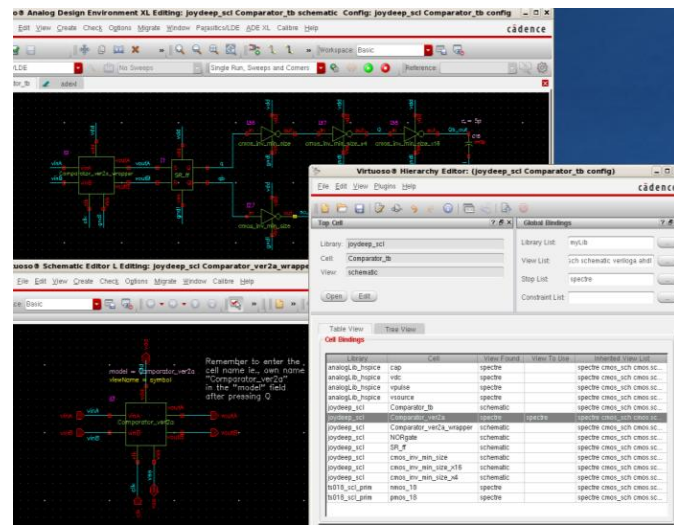

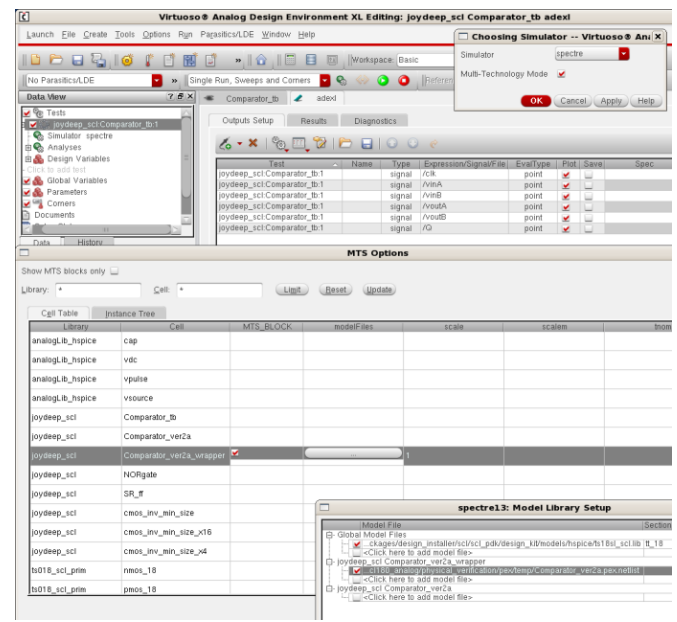

**Figure 4. Example screenshots showing creation and instantiation of wrapper cell view, setting up of Hierarchy Editor for config, ADE-XL window, MTS Options form, and Model Library Setup form settings.**

## 5. MONTE-CARLO SIMULATION

Launch *ADE-XL* from the schematic window for which Monte Carlo simulation is to be run. As prompted, create a new *adexl* view after which the ADE-XL GUI opens [14]. In the *Tests* section, click to create a new test due to which an ADE child window opens. Here, ensure that the *Simulator* option is set to *Spectre*, and set the model file to *ts18sl_scl_mat.lib* (instead of the usual *ts18sl_scl.lib*). Then, set the simulations and outputs to be saved/plotted.

Now return back to the ADE-XL window. Select *Monte Carlo Sampling* in the drop-down list. Click on Settings and set the fields as required. Then run the simulation.

## 6. I/O LIBRARY

*CIO150* is the 1.8V I/O pad library (*CIO250* being the 3.3V one) of SCL, that will be used here for demonstrating I/O ring.

### 6.1 Different I/O pads

Different types of I/O pads are available in the *CIO150* library which is present within the home directory. From this, the pads required commonly are listed below (and shown in Fig. 5):

- pc3d00 (used for analog I/O connection)
- pv0a (used for ground connection)
- pv0i  (used for ground connection)
- pv0c (used for ground connection)
- pvda (used for power connection)
- pvdi (used for power connection)
- pvdc (used for power connection)

Apart from these, there are layouts of many other types of pads available in the I/O library [15]. Together with I/O pads, we will need four corner cells and suitable filler cells to make the layout of the ring. The layout of fillers (e.g., *pfeed00010*,..) of various widths are present in *CIO150*, while the corner cells (comprised of *pfrelf* and *dummy_corner* cells) should be obtained from SCL (e.g., from their *io_template* library).

Fig. 5 depicts the use of various pads and cells to compose the I/O ring. The ring has four supply lines (i.e., VDD, VDDO, VSS and VSSO required for the circuits within the I/O pads). Further, as evident from the figure, the supplies from the *pvdi* and *pv0i* pads (i.e., VDD and VSS) may be used for the core circuits as well.

### 6.2 Composing the I/O Ring

For the sake of demonstration, let's make a new library named "Inverter_with_IO" to create a sample I/O ring. Copy all cells that you will need (e.g., *pvdi*, *pv0i*, etc.) from *CIO150* and *io_template* to your "Inverter_with_IO" library.

#### 6.2.1 Making Symbols of the Pads

After copying the layouts of the individual cells, we have to make corresponding symbols for each. The information about port names of individual pads are in the file *tsl18cio150.cdl* provided by SCL (shown in Fig. 6). To make the schematic and symbol views for individual pads, click on the respective cell (e.g., "pvdi" from the library "Inverter with IO"), then do

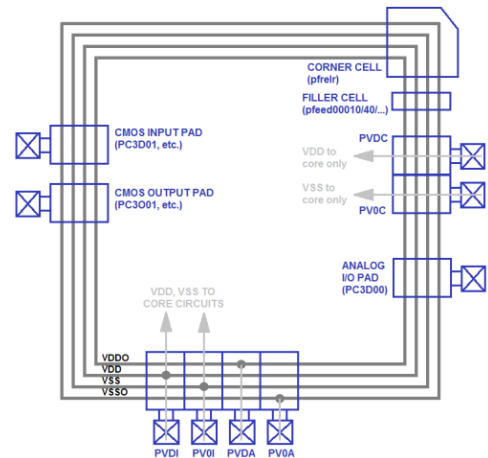

**Figure 5.  Depiction of general constituent cells of I/O ring with usage.**

*File→New→Cell View* from Library Manager. Like this, make symbols for all your pads (as depicted in example in Fig. 6).

In this regard, it should be mentioned that symbols are not required to be made for the corner and filler cells.

#### 6.2.2  Making the Ring

Using the symbols created earlier, compose the schematic of the ring. Ensure all the individual VDD, VDDO, VSS, VSSO are connected by means of proper labeling (as seen in Fig. 7).
*Note:* Put pins for at least a pair of Vdd & ground (say, VDD & VSS) in the schematic (as this is needed by LVS).

Further, create the corresponding layout (i.e., the I/O ring) for this schematic by instantiation of layouts of the required cells (including the fillers and corner cells). Consequently, the completed layout will be something like that shown in Fig. 7.
*Note:* Make sure that the individual cells are perfectly abutted, i.e., neither overlapped nor separated. It should give no DRC error and warning (apart from "projection parallel" warning)! Put labels for the Vdd & ground (here, VDD & VSS) as used in the schematic on the corresponding pads.

### 6.3  LVS of the I/O Ring

After DRC cleaning the layout, go for its LVS. To check LVS, we need to generate CDL netlist of the schematic of the ring (using same procedure as described before). Let the file name be "ring.cdl" for example.

Modify the generated CDL file in the following manner:
(i) Comment all the sub circuit definitions of the cells used e.g.

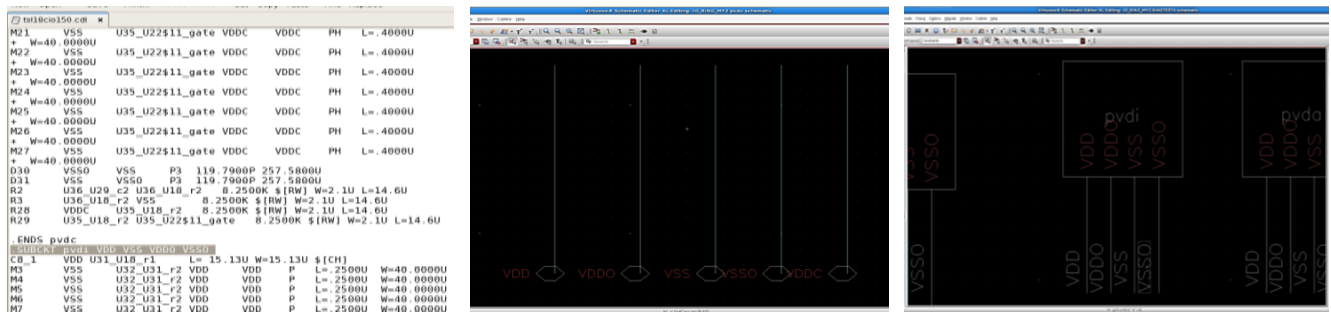

**Figure 6.  Example screenshots showing ts018150.cdl file, schematic prepared for pvdi, and an instantiated symbol of pvdi cell.**

pvda, pvdi, etc. using '*' symbol (except the main SUBCKT which defines the whole ring). E.g., for pvda:

```
*.SUBCKT pvda VDD VDDO VSS VSSO
*.PININFO VDD:B VDDO:B VSS:B VSSO:B
*.ENDS
```

(ii) Arrange/edit all the port names associated with the individual pad cell instances, defined inside the main SUBCKT named "ring" in the order as defined in the *tsl18cio150.cdl*. E.g., in the generated CDL, it may be like:

```
XI0 VDD VDDO VSS VSSO / pvda
```

But, in tsl18cio150.cdl, for pvda cell:

```
.SUBCKT pvda VDDO VDD VSS VSSO
```

So, appropriately edit the order/name of the ports like below:

```
XI0 VDDO VDD VSS VSSO / pvda
```

(iii) After editing CDL file, launch Calibre LVS and include the files *tsl18cio150.cdl* (first) and *ring.cdl* (second). The order is important (as all the devices/pads used in ring.cdl are defined in tsl18cio150.cdl file).

(iv) Now run LVS. If every step mentioned are followed correctly, the ring should be LVS clean.

### 6.4 LVS of Core Together with I/O Ring

For this, make a schematic instantiating (and interconnecting) your core and I/O ring blocks. Correspondingly, create the layout, containing the core and ring layouts, and route to the pads as required. Also, note the following:

(i) For analog I/O pad *pc3d00*, connect from core to the metal layer labeled as "PADR".
(ii) For the *pv0i*, *pvdc*, etc. power pads, connect using Metal-2 to the corresponding labels on the pads.
(iii) Using labeling layer for TOP_M, put labels on all the pads of your I/O ring as necessary.

Next, perform DRC and LVS of this whole layout following the same steps as mentioned in previous section. *Note:* if you are using CMOS input pad *pc3d01* or the output pad *pc3o01*, Calibre LVS will throw the following type of error:

```
property w not found on I9/RM30 (R)
```
(here I9 is an instance of pc3o01)

To clear this LVS error, comment the following lines within your LVS header file, as shown below:

```
// Select w/l property check for res/cap devices
//#define ANALOG
```

### 6.5 PEX and Simulation of Core Together with I/O Ring

Perform the PEX of the whole layout, for which the *Inputs→Netlists* are to be loaded in the same way as done for LVS.

Then, you can run post-layout simulation of this extracted netlist. Here, you will need to include the following extra sections from the model file ts18sl_scl.lib (required for the devices present within the pads, like NHV MOS, DPH diode, NWCAPH2T): "tt_hv", "diodes", "acc_typ".

### 6.6 Schematic Simulation of Core Together with I/O Ring

Circuit schematics of the I/O pads are not available and only netlist definition of these have been provided by the foundry in the file *tsl18cio150.cdl*. Thus, to simulate the total schematic of core jointly with its I/O ring, the method involving *config* view and *MTS* (as also discussed previously) is required. This is done using the steps as described below.

(i) Create a *symbol* view for the schematic that comprises of the I/O ring and core. Instantiate this symbol in a testbench schematic, and invoke ADE-XL. Set *Spectre* as the simulator, and enable the MTS option (steps for this has been discussed before). *Note:* You need to edit CDF and create *spectre* views for each of the I/O pads used in your circuit (using procedure as delineated in earlier section). For editing the CDF, the *termOrder* is required to be filled as defined for that particular pad in *tsl18cio150.cdl*.

(ii) Create and define a corresponding *config* view as needed: we need to use netlist definition for I/O pads (hence, *spectre* view to be used), and schematic definition of core circuits.

(iii) In *MTS Options* form, click to check and hence, enable MTS for the constituent core blocks. Set *scale* to 1e-6 for these blocks. However, keep the *scale* at 1 in ADE *Simulator Options* form.

(iv) As the provided *tsl18cio150.cdl* netlist is for LVS etc. purposes, so the devices are defined using their LVS names (refer to SCL Design Rules manual [16] for details). Hence, manually replace those names in this file with corresponding SPICE names (for all of the I/O pads used). E.g., the LVS name of 1.8V PMOS is "P" while its SPICE name is "P18".

(v) In the *Model Library Setup* form of ADE, include the path to the edited *tsl18cio150.cdl* file.

With these settings, you are ready to simulate in ADE-XL.

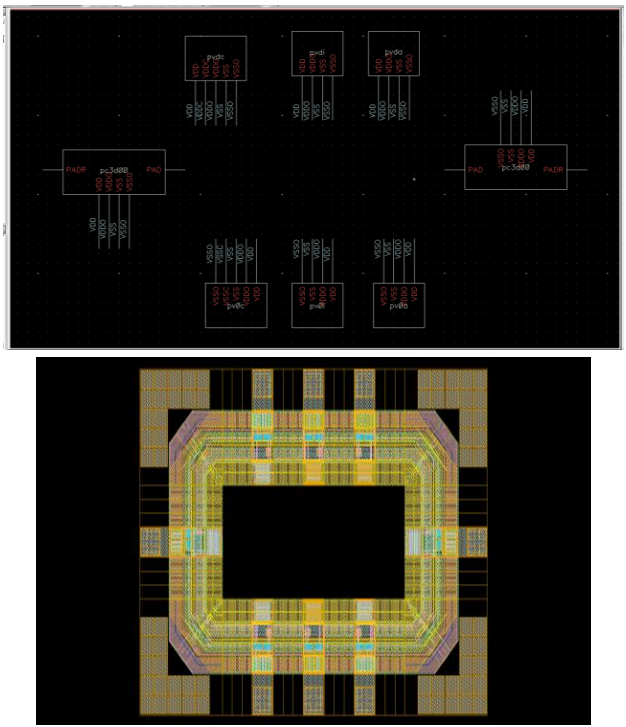

**Figure 7. Example showing creation of I/O ring schematic and layout.**

# 7. FINAL STEPS TOWARDS TAPE-OUT

Once the layout of the core together with its I/O ring is completed and have been subjected to DRC, LVS and Antenna checks, you are ready to embark on the final tasks for arriving at the full-chip GDS file. The following sections describe such steps in detail that are imperative for a successful tape-out.

## 7.1 Seal Ring

The chip layout should be enclosed within a metal seal ring, as depicted in the sample layout in Fig. 8. A seal ring block is supplied by SCL in Top_TS18SL_EXTERNALS_10.10.gds with cell name TS18_SealRing_SL_4M1L_a0. The GDS is required to be Streamed in (by invoking the *Stream in* tool from Virtuoso CIW window using *File → Import → stream*). The generated ring should be appropriately resized and then instantiated in the layout surrounding the I/O ring.

The designer is advised to refer to SCL's DRC manual [16] for seal ring rules. Also, the seal ring must be connected to either VSS (from *pvdi*) or VSSO (from *pvda*) of the I/O ring.

## 7.2 Silicon Number

The designer needs to contact SCL PDK team in advance for assigning a unique *Silicon Number* for the intended tape-out. A GDS file Ext_Str_TSL_Shuttle_4M1L.gds having the silicon number layout block should also be obtained. Stream in this GDS in a new library, and use the "SiNumberLargeParent" cell in your layout. Edit individual characters of this default silicon number block by changing their property to obtain the appropriate number as provided by SCL for your design. Make sure that all the characters are fully enclosed by the outer blue layer. Also, this block should be placed in the bottom-left corner of the layout as seen in Fig. 8.

## 7.3 Dummy Fill

Before dummy filling, you might consider filling any vacant space with decoupling capacitors (MOS caps between VDD and GND) if required. After completion of all cell placement and routing activities, dummy insertion is be done for active (AA), poly (GC) and metals as per the flow steps described below. For particular analog blocks that might be sensitive to dummies, handcrafted Dummy Filling might be done on the top level of the blocks which should then be covered with NODUMMY layers.

(i) Stream out (using *Stream out* tool from Virtuoso CIW window using *File → Export → stream*) to create the GDS file (e.g., FINAL_LAY.gds) for the final DRC cleaned (normal i.e., block-level) layout (say FINAL_LAY, positioned at (0,0) coordinates) where you want to fill the dummy metal layers.
(ii) Get the *DUMMYFILL_TS18SL_SCL_CALIBRE* and *DUMMY.header* files from the following path and copy in home folder.

```
~/scl_pdk/design_kit/drclvs_rsf/DUMMY.header
~/scl_pdk/design_kit/drclvs_rsf/DUMMYFILL_TS18
SL_SCL_CALIBRE
```

(iii) Edit this *DUMMY.header* file providing paths of the input GDS, dummy filled output GDS, etc., as shown in the example

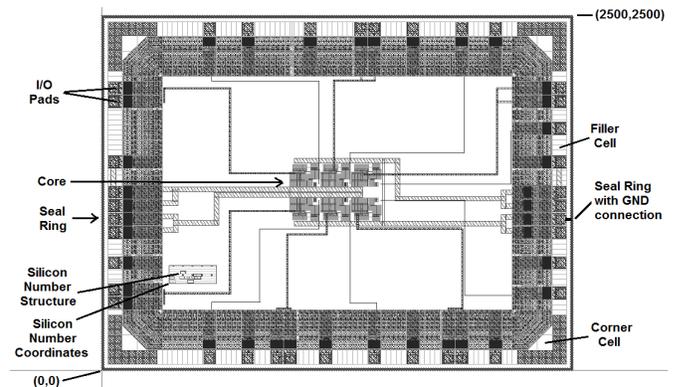

**Figure 8.** Sample layout showing core, I/O ring, seal ring, and silicon number structure.

given in Fig. 9.
(iv) To generate (say, having name) FINAL_LAY_dummy.gds file having the dummies, run the following two commands:

```
source .bashrc (or equivalent command that you use)
calibre -drc ./DUMMY.header
```

(v) Create a new library (with TSL tech file) and do a Stream in of the generated FINAL_LAY_dummy.gds from CIW. This will create the corresponding layout with dummy layers.
(vi) Instantiate this dummy filled layout in your original layout FINAL_LAY at origin (0,0) using *properties→Origin X, Y*. This will produce the dummy filled full-chip layout version.

## 7.4 Full-Chip DRC

To enable checking of DRC for the full-chip layout, edit the *DRC.header* file by uncommenting the line "#define CHIP" while commenting the line "#define BLOCK" as shown below (so that DRC would check for 421 rules instead of 356 rules in block level), and then run DRC using same process as before:

```
#define CHIP
//#define BLOCK
```

This DRC generates a results-summary file which (for the final layout) is required to be submitted to SCL with the GDS.

Antenna Check must also be done in the same way as DRC. The results-summary file from Antenna Check run is also required to be submitted to SCL.

## 7.5 GDS Generation

Generate GDSII file for the final full-chip layout by invoking *Stream out* tool from Virtuoso CIW. During this translation, a *streamout.log* file gets generated by default. Along with this, a summary file can also be generated by providing a suitable name (without extension) in the field *Stream Out Summary* (in Option settings of *Stream out* tool). By inspecting both the files, we can find out the names of the layers as used in the layout. List out these layer names in a text file. This should be submitted to SCL together with the GDSII.

## 7.6 CRC Checksum

The CRC checksum consists of a pair of numbers that is used by the foundry to verify the propriety of the GDS file received

```
LAYOUT PATH "/mtech/sanjay/scl180_analog/FINAL_LAY.gds"
            LAYOUT PRIMARY "FINAL_LAY"
            LAYOUT SYSTEM GDSII

            // Output Database/report
            DRC SUMMARY REPORT
"FINAL_LAY_dummy_generation_drc_report" REPLACE HIER
            DRC RESULTS DATABASE  "FINAL_LAY_dummy.gds"
GDSII

            DRC MAXIMUM RESULTS ALL
            DRC MAXIMUM VERTEX ALL
            DRC CELL NAME YES CELL SPACE XFORM
            // User defined options.
            //--------------------
            DRC ICSTATION YES

// Select Metal Options
//#define 2LM
//#define 3LM
#define 4LM
//#define 5LM
//#define 6LM
//#define 4M1T

// Select Generated Layers
#define AA_gen
#define GC_gen
#define M1_gen
#define M2_gen
#define M3_gen
#define M4_gen
#define M5_gen
#define MT_gen

// Select Density Check
//#define DENSITY_CHECKS

INCLUDE /mtech/sanjay/scl180_analog/DUMMYFILL_TS18SL_SCL_CALIBRE
```

**Figure 9.** Example showing how to edit the *DUMMY.header* file.

from the user. Download the "mosiscrc.c" file from [17] and copy it to a new folder together with your GDSII file. Open a terminal there and enter following command:

```
gcc mosiscrc.c -o mosiscrc
./mosiscrc -b <your_GDSII_file_name.gds>
```

This will display the checksum for the concerned GDSII file.

**7.7 Sending Files to SCL**

Fill up *SCL tapeout submission form* doc file with required details pertaining to the layout, e.g., customer information, design environment/version used, GDSII file name/size, die size/coordinates, types of core device and I/O library used, checksum value, etc. This is to be submitted together with the GDSII file, reports/logs of DRC, Antenna, Stream out; and lists of device types and layers used in the design. The list of devices is a text file containing names of devices used in both the core and I/O ring of the design, like *pmos_18*, *nmos_18*, *cmim_sq*, *pdio_sal*, etc.

## 8. CONCLUSION

The ESDM industry is of strategic importance to India from the perspectives of economic growth, enhancement of high-technology base, employment generation, and national security concerns. Consequently, the Govt. of India attributes high priority to domestic electronic hardware manufacturing up to the chip level. A promising step in this regard is the foundry of SCL at Chandigarh, that has been commissioned with 180nm CMOS fabrication facility a few years back. In order to realize designed VLSI circuits, researchers from Indian academia and R&D centres should take advantage of the Multi Project Wafer runs of SCL that promises to be a relatively cost-effective option in contrast to commercial foundries abroad.

In order to effectively design and tape-out using SCL's technology, various information are required during the entire process. This paper is aimed at providing all such inputs that might be generally needed in the process of taping out Analog VLSI circuit designs to SCL, and should be immensely helpful for designers intending to submit their designs to SCL.

### APPENDIX

Information on environment/version of EDA tools as utilized:
*Cadence*: Virtuoso version IC6.1.5.500.17
*Mentor Graphics*: Calibre v2013.2_18.13
*Synopsys*: HSPICE version J-2014.09-2

The author wishes to state that the best effort has been put to avoid cases of error or missing information in this tutorial. However, any such instance if found is completely inadvertent. The tool usage methods discussed here are not necessarily best case ones, and there may be more efficient ways for the same.

### ACKNOWLEDGMENT

The author would like to sincerely thank Mr. H. S. Jattana, Mr. Uday P. Khambete, and Mr. Ashutosh Yadav of SCL, Chandigarh for information and support received during the course of interactions. The author would like to express deep gratitude to Prof. Pradip Mandal of E&ECE, IIT Kharagpur for all the helpful advice and guidance. Appreciations are also due to Dr. Samiran Dam and Mr. Koustav Roy (formerly with AVDL, IIT Kharagpur) for their documentation efforts related to the usage of SCL process; and to Prof. T. K. Bhattacharyya (of E&ECE and Professor-in-charge, AVDL), Prof. Mrigank Sharad (of E&ECE), Dr. Nijwm Wary (formerly with AVDL), and Mr. Indranil Som and Mr. Hrishikesh Sarkar (of AVDL) for their various helpful efforts and kind support.